\documentclass[aps,prd,superscriptaddress,nofootinbib,floatfix,twocolumn]{revtex4-2}

\pdfoutput=1
\usepackage{graphicx}
\usepackage{amsmath}
\usepackage{bm}
\usepackage[small,bf]{caption}
\usepackage[subrefformat=parens]{subcaption}
\usepackage{comment}

\usepackage{tabularx}
\usepackage{array}
\newcolumntype{R}{>{$}r<{$}}  

\usepackage{booktabs}

\newcommand{\beq}{\begin{equation}}
\newcommand{\eeq}{\end{equation}}
\newcommand{\beqa}{\begin{eqnarray}}
\newcommand{\eeqa}{\end{eqnarray}}
\newcommand{\bpr}{\begin{problem}}
\newcommand{\epr}{\end{problem}}
\newcommand{\bcent}{\begin{center}}
\newcommand{\ecent}{\end{center}}
\newcommand{\bfig}{\begin{figure}}
\newcommand{\efig}{\end{figure}}
\newcommand{\bpc}{\begin{picture}}
\newcommand{\epc}{\end{picture}}
\newcommand{\barr}{\begin{array}}
\newcommand{\earr}{\end{array}}
\newcommand{\bitm}{\begin{itemize}}
\newcommand{\eitm}{\end{itemize}}
\newcommand{\bright}{\begin{flushright}}
\newcommand{\eright}{\end{flushright}}
\newcommand{\bminip}{\begin{minipage}}
\newcommand{\eminip}{\end{minipage}}
\newcommand{\btab}{\begin{tabular}}
\newcommand{\etab}{\end{tabular}}

\newcommand{\hiroshima}{Graduate School of Advanced Science and Engineering, Hiroshima University, Kagamiyama, Higashi-Hiroshima, Hiroshima 739-8526, Japan}
\newcommand{\QUP}{International Center for Quantum-field Measurement Systems
for Studies of the Universe and Particles (QUP), KEK, Tsukuba, Ibaraki 305-0801, Japan}

\newcommand{\dd}{\mathrm{d}}
\newcommand{\dv}[2]{\frac{\dd #1}{\dd #2}}

\begin{document}
\title{Earth-lens telescope for distant axion-like particle sources \\ with stimulated backward reflection}

\author{Taiyo Nakamura}\affiliation{\hiroshima}
\author{Kensuke Homma*}\affiliation{\hiroshima}\affiliation{\QUP}

\begingroup
\renewcommand\thefootnote{\fnsymbol{footnote}} 
\footnotetext[1]{Corresponding author}
\endgroup

\date{August 18, 2025}

\begin{abstract}
We propose a novel telescope concept based on Earth’s gravitational lensing effect, optimized for the detection of distant dark matter sources, particularly axion-like particles (ALPs). When a unidirectional flux of dark matter passes through Earth at sufficiently high velocity, gravitational lensing can concentrate the flux at a distant focal region in space. Our method combines this lensing effect with stimulated backward reflection (SBR), arising from ALP decays that are induced by directing a coherent electromagnetic beam toward the focal point.
The aim of this work is to numerically analyze the structure of the focal region and to develop a framework for estimating the sensitivity to ALP--photon coupling via this mechanism. Numerical calculations show that, assuming an average ALP velocity of 520\,km/s---as suggested by the observed stellar stream S1---the focal region extends from $9 \times 10^9$\,m to $1.4 \times 10^{10}$\,m, with peak density near $9.6 \times 10^9$\,m. For a conservative point-like ALP source located approximately 8\,kpc from the solar system, based on the S1 stream, the estimated sensitivity in the eV mass range reaches $g/M = \mathcal{O}(10^{-22})\,\mathrm{GeV}^{-1}$.
This concept thus opens a path toward a general-purpose, space-based ALP observatory that could, in principle, detect more distant sources---well beyond $\mathcal{O}(10)\,\mathrm{kpc}$---provided that ALP--photon coupling is sufficiently strong, that is, $M \ll M_\mathrm{Planck}$.
\end{abstract}

\maketitle

\section{Introduction}
Although there is no inherent mechanism in quantum chromodynamics (QCD) to preserve CP symmetry, CP symmetry is nevertheless observed to be extremely well preserved within the QCD sector. To explain this unnaturally preserved symmetry, a global U(1) symmetry known as the Peccei–Quinn (PQ) symmetry was proposed \cite{PQ_symmetry_PRL, PQ_symmertry_PRD}. The spontaneous breaking of this symmetry gives rise to a pseudo–Nambu Goldstone boson with non-zero mass, known as the axion \cite{Wilczek_axion, Weinberg_axion}.
When the PQ symmetry is broken at an energy scale higher than that of the electroweak phase transition, the coupling between the axion and Standard Model particles becomes extremely weak. Because of this property, the axion is considered a viable candidate for cold dark matter.
Regarding the theoretical predictions of the mass–coupling region between the axion and ordinary matter, the astrophobic axion model suggests that the axion decay constant $f_a$ can be constrained to $f_a = 10^{6-7} \,\mathrm{GeV}$ \cite{Astrophobic_axion1, Astrophobic_axion2}. If the anomalous cooling of stars is interpreted as being caused by the axion, then the axion mass at that time is given by $m_a = 5.70 \left( \frac{10^9\,\mathrm{GeV}}{f_a} \right) \,\mathrm{meV}$ \cite{Stellar_cooling_anomary}.
Under the aforementioned constraint on the decay constant, the axion mass falls in the range of $\mathcal{O}(10^{0-1})\,\mathrm{eV}$.
Similarly, for example, an axion-like particle (ALP) model that allows cosmic inflation also provides predictions 
including the $\mathcal{O}(10^{0-1})\,\mathrm{eV}$ mass range \cite{ALP_miracle2018}.
In particular, hints of the line spectra in the eV mass range as discussed in Ref.~\cite{Yin:2024lla}
stimulates deeper searches for ALPs specifically in the eV mass range.
These considerations indicate that the search for weakly photon-coupled particles, such as the axion and ALPs in the eV mass range, is of significant interest.

\begin{figure}[h]
\centering
\includegraphics[width=0.5\textwidth]{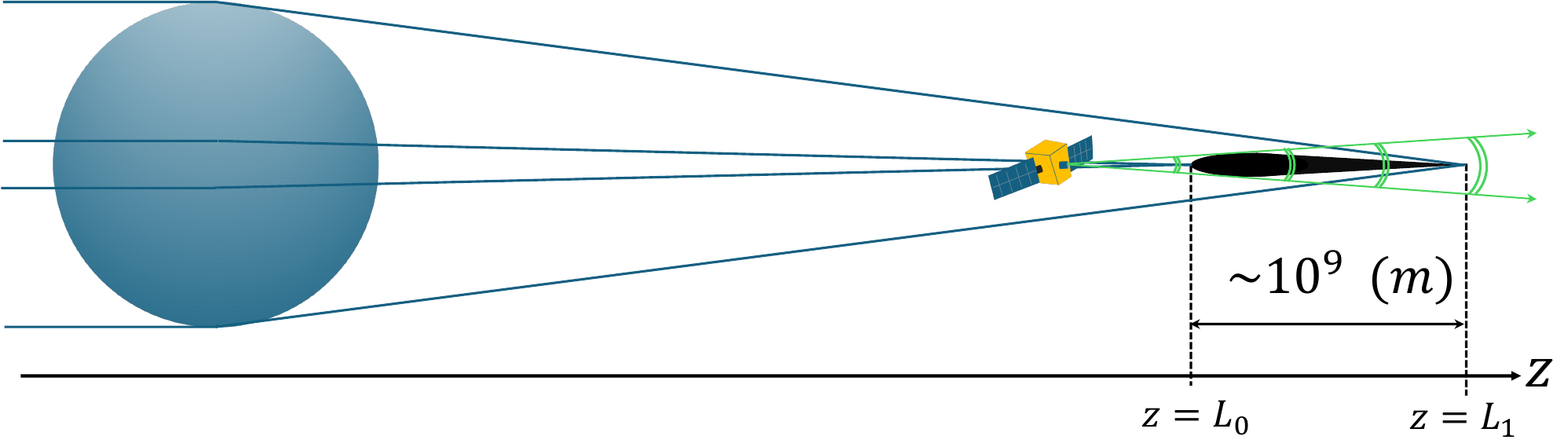}
\caption{
Search geometry of an axion-like particle (ALP) lensing object using the Stimulated Backward Reflection (SBR) method from a space probe (yellow).
The blue circle denotes the Earth, while the blue lines indicate the trajectories of individual ALPs incident from the left. The black triangular structure with trailing tails illustrates the concentrated distribution of dark matter at the focal region. The localized ALP flux is probed by a spherical wave of a coherent electromagnetic field to induce decay of the ALPs.
$L_0$ denotes the emission point of the inducing field, and $L_1$ represents the effective terminal point beyond which the field strength becomes negligible. The total probing distance, $L_1 - L_0$, extends to approximately $10^9$~m.
}
\label{Fig1}
\end{figure}

Gravitational lensing by the Earth focuses a unidirectional dark matter (ALP) flux with incident velocity 220~km/s at a distance of $\mathcal{O}(10^9)$~m from the Earth \cite{Przeau_2015}.
In ALP searches, the conventional method~\cite{Sikivie:1983ip} uses large magnets 
based on the inverse Primakoff process.
A magnetic field provides virtual photons that convert an ALP into a real photon~\cite{CAST:2017uph, ADMX:2024xbv}. 
However, it is likely difficult to carry magnets to a distant focal point and to supply the required field strength over the required volume. In contrast, as illustrated in Fig.\ref{Fig1}, the stimulated backward reflection method (SBR) uses real photon fields such as a laser field to scan space \cite{Homma2024}. The effective volume of the probe field does not depend on the physical dimensions of the light source.
In other words, a sufficiently strong inducing field enables the exploration of distant regions, such as the focal point of the ALP flux.
An additional advantage of SBR is that the signal photons return to the emission point of the inducing field. The inducing field propagates as a spherical wave over long distances. The coherency of the field stimulates the decay of an ALP into two photons on its wavefront. In this configuration, the momentum and polarization states of one of the two decay photons can be selected to match the local momentum and polarization of the inducing field on the spherical wavefront. Consequently, the other photon must be emitted in the direction that satisfies energy–momentum conservation, which requires that the photon pair be emitted orthogonally to the wavefront. As a result, the wavefront effectively acts as a mirror, and the signal photon propagates in the direction opposite to that of the inducing field. Like light reflected from a parabolic mirror, the signal photon is reflected backward to the emission point of the inducing field.

Since the focal region lies beyond the Earth's atmosphere, it is natural to position 
the emission point of the inducing field in space in order to suppress background reflections arising 
from Rayleigh scattering. To maximize overlap with the densest portion of the ALP flux, 
it is preferable to locate the light source in proximity to the focal region, 
thereby directing the most intense part of the inducing field into the most concentrated 
region of the flux. Accordingly, deploying a space probe equipped with a coherent light source 
to the focal region and emitting the inducing field directly from the probe constitutes 
an effective strategy~\cite{Homma2024}.
However, to thoroughly probe the concentrated ALP flux and to configure the inducing field 
as a near-ideal spherical wave, it is essential to determine an optimal observational position 
for the space probe based on the detailed structure of the gravitational lensing object. 
Moreover, in the previous work~\cite{Homma2024}, the individual ALP velocity vectors 
around the focal region were not calculated at all 
despite its importance for evaluating the signal collection efficiency—
a crucial factor in accurately estimating the sensitivity to the ALP-photon coupling
Therefore, this paper addresses these essential issues through detailed numerical calculations.

In this paper, we discuss a conservative evaluation on the effective incident ALP density 
to the Earth by assuming the S1 stream from the Cygnus constellation~\cite{OHareS1stream} in section 2.
We then provide the numerical method employed for this purpose in section 3.
In section 4 we show the simulated result and discuss the resulting spatial structure of the focused ALP flux.
In section 5 we evaluate the acceptance factor of reflected signal photons based on the velocity distributions of 
the simulated ALP flux in individual segments over the propagation region of the inducing field.
In section 6 we provide the more accurate sensitivity projection based on the numerically calculated focused-flux and the acceptance factor of reflected signal photons by taking non-relativistic motion of ALP flux due to the lensing effect into account. 
Section 7 concludes with a summary of the proposed Earth-lens telescope for distant ALP sources, 
while section 8 presents future prospects along with supporting discussions. 

\section{A model for incident ALP density evaluation from distant ALP-emitting objects}
Reference~\cite{Przeau_2015} presents analytical expressions for both the focal length and the magnification of the axion-like particle (ALP) density along the incident axis, derived from geodesic trajectories under the Schwarzschild metric, in the case where a unidirectional ALP stream traverses the Earth as illustrated in Fig.\ref{calc geometory}.  
According to these formulations, for particles skimming the Earth---i.e., those with trajectories outside its surface---the magnification is directly proportional to the impact parameter \( b \), implying the potential for larger magnification relative to trajectories that penetrate the Earth's interior. However, the corresponding focal length increases as \( b^2 \), thereby reducing the ALP density in the focal region with increasing \( b \), which in turn diminishes the overall effectiveness of the gravitational focusing.  
Conversely, for particles that pass through the Earth’s interior, this focusing effect remains significant, as the focal length does not exhibit the same quadratic growth.  
In this study, we numerically evaluate the ALP density in the gravitational focal region, adopting the S1 stream originating from the Cygnus constellation~\cite{OHareS1stream} as a concrete astrophysical candidate for a unidirectional ALP flux traversing the Earth’s interior.  
The rationale for selecting the S1 stream over the Galactic ALP halo lies in the intrinsic anisotropy of the former. While ALP halos may appear stationary in the Galactic rest frame, they are generally considered to be virialized, having attained a quasi-equilibrium state through gravitational interactions within the Galactic disk. As a result, their constituent particles exhibit thermal velocities on the order of the Galactic rotation speed. Thus, despite the Solar System's motion at approximately 220~km/s relative to the Galactic center, the associated thermal dispersion of the ALPs prevents the assumption of a well-collimated, unidirectional flux incident on the Earth.  
By contrast, dark matter associated with the observed S1 stream---an astrophysical structure presumed to be the remnant of a disrupted dwarf galaxy---passes through the Solar System with a bulk velocity of approximately 520~km/s~\cite{OHareS1stream}. Owing to the considerable distance to the progenitor system, exceeding 8~kpc, the ALP flux originating from the S1 stream impinges upon the Earth as a plane-parallel beam, exhibiting minimal angular divergence on the order of \( \mathcal{O}(10^{-13}) \) radians~\cite{Homma2024}. This exceptional collimation renders the flux highly susceptible to gravitational lensing effects, thereby enabling substantial concentration in the focal region.

One of the authors of the present study previously calculated the effective average density of axion-like particles (ALPs) incident on the Earth's gravitational lens, under the assumption that an ALP flux is streaming along the observed S1 stream in the earlier study~\cite{Homma2024}. This density estimate was then used to evaluate the sensitivity to ALP–photon coupling. However, the S1 stream serves merely as a representative example; the same framework is applicable to unknown, distant ALP sources as well.

As the simplest and most conservative estimate of the incoming ALP flux, we assume that the source emits ALPs isotropically, in analogy with distant stars emitting light. In this case, the farther the ALP source is from the solar system—where the space probe is located—the smaller the solid angle subtended by the detector onboard the probe, leading to a reduced observed flux. Nevertheless, this narrowing of the solid angle also enables the identification of the source direction, effectively transforming the system comprising the planetary lens and the probe into a new type of telescope.
Moreover, the planetary lens effect causes a convergence of the ALP flux along the line of sight to the source, introducing spatial anisotropy in the flux that can be exploited for directional ALP astronomy. In what follows, we formalize the effective average density of the ALP flux incident on the Earth's gravitational lens, assuming such a distant ALP-emitting object.

We thus introduce a conservative model on the ALP density $\rho(v)$ at the incident plane in front of the Earth lens for a given incident stream velocity $v$.
Suppose an ALP-emitting object such as a dwarf galaxy with the total energy $E(t)$ which
isotropically emits an ALP flux as a function of time $t$ in the process of infall with $E(0) = M_0$ corresponding to the total mass of the object.
For a short time period compared to the infall time scale $T_{in}$ of the object,
the emission rate can be approximated with a differential equation
$dE = -\tau^{-1} E dt$ as a result of emission
of ALP above the escape velocity $v_{esc}$.
We assume $\tau \sim T_{in}$ because the age of the Universe may be similar to
the infall time scale.
In this case the emission rate $r_e$ observed at the incident plane of the Earth lens
with distance $D$ to the source is approximated as
\begin{equation}
r_e \sim \tau^{-1} M_0 \frac{\pi R^2_E}{4\pi D^2}
\end{equation}
where the radius of the Earth $R_E$, $M_0 = 10^{10} M_{\odot}$, and $\tau = 10^{10}$~yr were used for the case of the S1-stream~\cite{Homma2024}.
Given the incident rate, the local energy density flowing into a cylindrical
volume with the base area corresponding to the Earth's cross section
in short time duration $\Delta t$ is expressed as
\begin{equation}
  \rho(v) = \frac{r_e \Delta t}{v \Delta t \pi R^2_E} = \frac{\tau^{-1} M_0}{4\pi D^2 v}.  
\end{equation}
We then assume the velocity distribution of ALP follows Gaussian distribution and
the probe receives an ALP flux within the narrow solid angle.
In this case the ALP energy density at the incident region around the Earth
is expressed with an effectively one-dimensional velocity dispersion as follows
\begin{equation}
\label{eq_rhos}
\rho_s  = \int_{v_{esc}-5\sigma_v}^{v_{esc}+5\sigma_v} dv
\frac{\rho(v)}{\sqrt{2\pi\sigma^2_v}} \exp\left(-\frac{(v-v_{esc})^2}{2\sigma^2_v}\right)
\end{equation}
where $\rho(v)$ is integrated with the Gaussian weight in the case of previous study~\cite{Homma2024} assuming
the escape velocity $v_{esc}=520$~km/s and the standard deviation $\sigma_v=46$~km/s
as discussed in~\cite{OHareS1stream}, resulting in
\begin{equation}\label{eq_frho}
 f_{\rho} = \rho_s/\rho_0 = 0.002    
\end{equation}
from Eq.(\ref{eq_rhos}), which is conservative compared to $f_{\rho} = 0.1$ with $\rho_0=0.5$ GeV/cm${}^3$ in \cite{OHareS1stream}.
We note that we can, in principle, distinguish the concentrated ALP flux from distant ALP-emitting objects based on the Doppler shift of the back reflected signal photons as indicated in the first line of Eq.(\ref{EqConservation})
from those caused by the local dark matter halo following the standard halo model with $v_{esc} < 300$~km/s.

\section{Method of the numerical calculation for the gravitational lensing effect}
In our calculation, the equations of motion for a Newtonian gravitational field generated 
by a uniformly dense Earth are reformulated into finite-difference form using the Euler method, and solved numerically.
To exploit the spherical symmetry of the problem, we adopt spherical coordinates $(r,\,\theta,\,\phi)$ for the computation. Assuming a constant density profile for Earth’s interior, the gravitational potential forms inside and outside 
the Earth of mass $M_E$ are given as follows:
\begin{equation}\label{potential}
\begin{aligned}
\text{External potential:} \quad U_{\mathrm{ext}}/m_a &= -\frac{GM_E}{r} \\
\text{Internal potential:} \quad U_{\mathrm{int}}/m_a &= -\frac{GM_E}{2R_E^3}(3R_E^2 - r^2),
\end{aligned}
\end{equation}
where $m_a$ is the mass of an ALP, $G$ is the gravitational constant, and $R_E$ is Earth’s radius.
Based on Eq.(\ref{potential}), the motion of the ALP follows the differential equations given below for the external and internal regions, respectively,
\begin{equation}\label{external newtonian}
\begin{aligned}
\text{External}& \text{ Newtonian equation:}\\
\ddot{r} &= r\dot{\theta}^2 - \frac{GM_E}{r^2} \\
\ddot{\theta} &= 0 \quad \Leftrightarrow \quad \dot{\theta} = \frac{j}{r^2}
\end{aligned}
\end{equation}
\begin{equation}\label{internal newtonian}
\begin{aligned}
\text{Internal}& \text{ Newtonian equation:} \\
\ddot{r} &= r\dot{\theta}^2 - \frac{GM_E}{R_E^3} r \\
\ddot{\theta} &= 0 \quad \Leftrightarrow \quad \dot{\theta} = \frac{j}{r^2}.
\end{aligned}
\end{equation}
Here, $j$ denotes the angular momentum per ALP mass.
As evident from the second lines of Eqs.(\ref{external newtonian}) and (\ref{internal newtonian}), angular momentum is conserved in the azimuthal ($\phi$) direction.
The ALP trajectory thus can be limited to the two-dimensional plane $(\phi = 0)$, i.e. $(r,\theta)$.
While Eq.(\ref{external newtonian}) admits an analytical solution, Eq.(\ref{internal newtonian}) does not. Therefore, to suppress computational errors at the surface of the Earth, we perform numerical integration for both regions.
The Euler form of the equations of motion is given below:
\begin{equation}\label{numerical equation}
\begin{aligned}
\text{External}& \text{ Euler equations:} \\
r_{n+1} &= r_n + \dot{r}_n\Delta t \\
\theta_{n+1} &= \theta_n + \frac{j}{r_n^2}\Delta t \\
\dot{r}_{n+1} &= \dot{r}_n + \Delta t \left( \frac{j^2}{r_n^3} - \frac{GM_E}{r_n^2} \right) \\
\dot{\theta}_{n+1} &= \frac{j}{r^2 _{n+1}} \\
\text{Internal}& \text{ Euler equations:} \\
r_{n+1} &= r_n +\dot{r}_n \Delta t \\
\theta_{n+1} &= \theta_n + \frac{j}{r_n^2}\Delta t \\
\dot{r}_{n+1} &= \dot{r}_n + \Delta t \left( \frac{j^2}{r_n^3} - \frac{GM_E}{R_E^3} r_n \right) \\
\dot{\theta}_{n+1} &= \frac{j}{r^2 _{n+1}}
\end{aligned}
\end{equation}
These difference equations enable the numerical integration of ALP trajectories across interior and exterior of 
the Earth under Newtonian gravity.

\begin{figure}[h]
\centering
\includegraphics[width=0.5\textwidth]{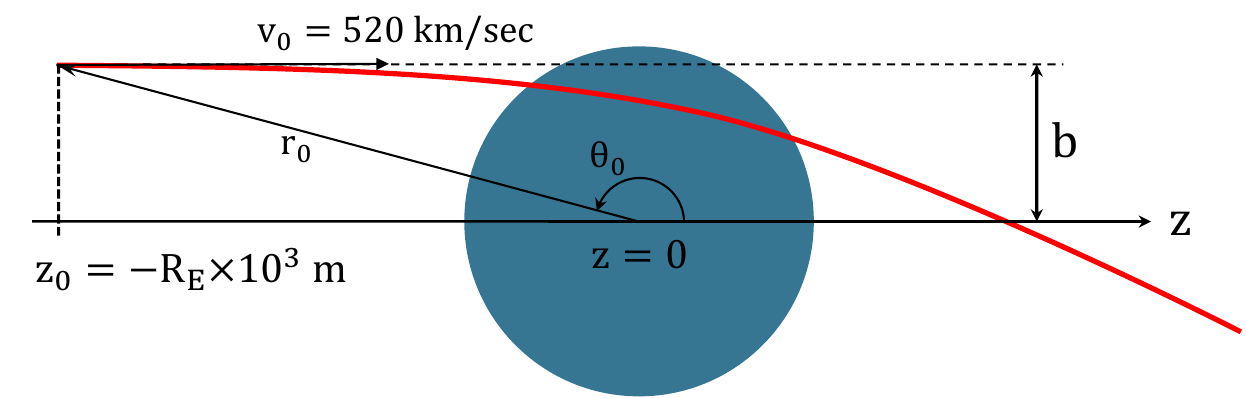}
\caption{
Geometry of a particle trajectory, the initial position, the initial velocity, and the impact parameter.
The ALP trajectory is shown with the red curve.
The initial position is located at a distance of $-1000R_E$ from the Earth’s center along the incident axis $z$, 
and the vertical offset by the impact parameter $b$.
The initial velocity is $v_0=520$~km/s, directed parallel to the incident axis.
}
\label{calc geometory}
\end{figure}

Next, we describe the initial conditions used in the following numerical calculation.
As shown in Fig. \ref{calc geometory}, the impact parameter $b$ is varied from $0$ to $R_E$ in 1000 equal steps. This gives us 1000 trajectories.
From the geometry in Fig. \ref{calc geometory}, the initial coordinate and velocity for a particle with a given $b$ are 
given by:
\begin{equation}
    \begin{aligned}
        (r_0,\,\theta_0) &= \left( \sqrt{z_0 ^2 + b^2},\, \theta_0 \right)\, \mathrm{m}\\
        (v_{r0},\, r_0 \dot{\theta})& = \left( v_0 \cos(\pi - \theta_0),\, v_0 \sin(\pi - \theta_0) \right)\, \mathrm{km/s}
    \end{aligned}
\end{equation}
Here, $z_0=-1000R_E$ and $v_0=520$ km/s, and these geometric relations are shown in Fig.\ref{calc geometory}.

\begin{figure}[h]
\centering
\includegraphics[width=0.53\textwidth]{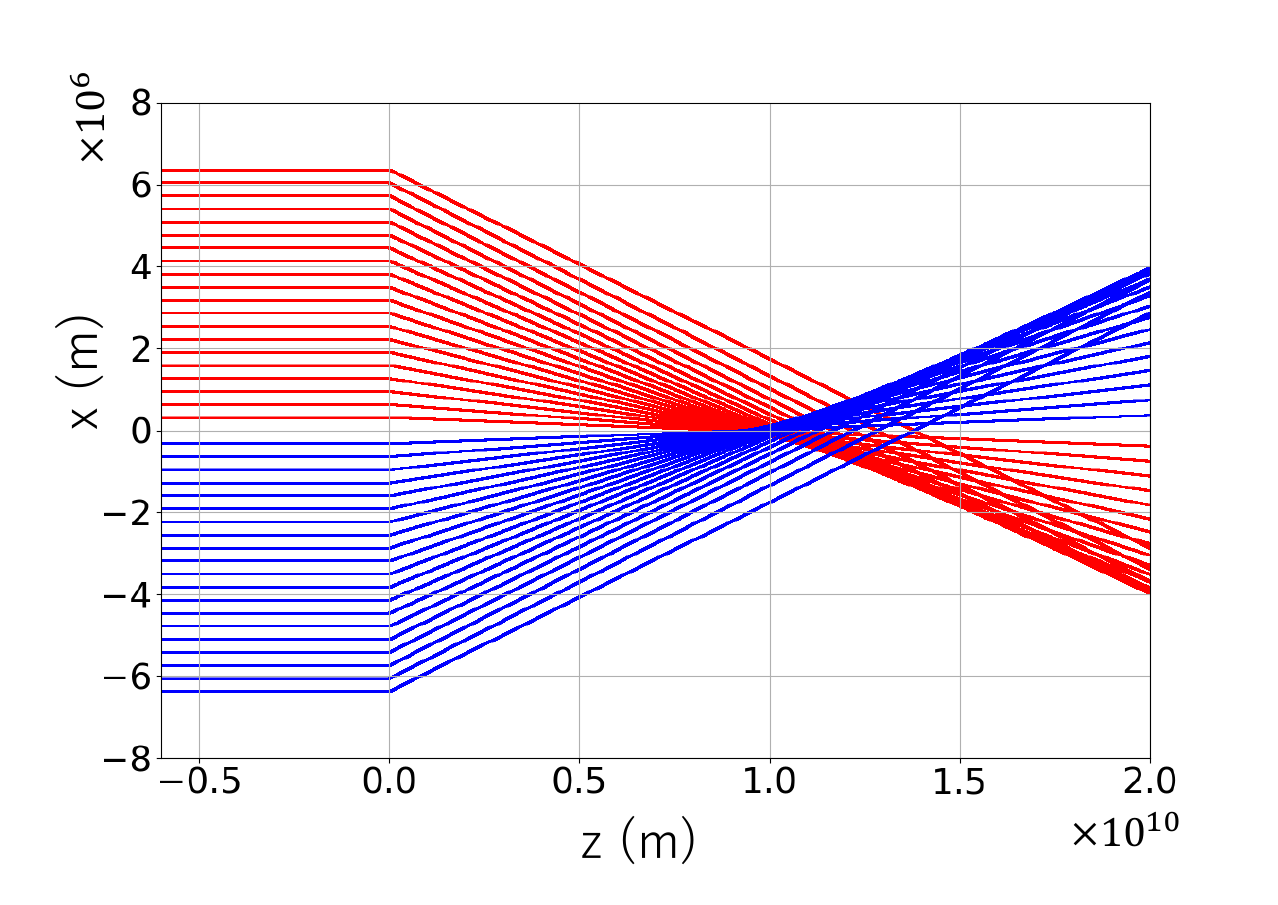}
\caption{
Deflection of particle trajectories.
Red curves correspond to trajectories with $b > 0$, while blue curves represent those with $b < 0$.
The focal points are distributed in the range $9.2 \times 10^9 \lesssim z \lesssim 1.4 \times 10^{10}$
for the incident velocity of $v_0=520$~km/s.}
\label{trajectories}
\end{figure}

Figure \ref{trajectories} shows the particle trajectories obtained from the numerical calculation, visualizing how the paths bend and intersect in the x-z plane in Cartesian coordinates.
The Earth is positioned at the origin, centered at $(x, z) = (0, 0)$.
It shows that the trajectories bend sharply near $z = 0$, while they remain nearly linear elsewhere. This indicates that the lensing effect is significant only in the vicinity of the Earth.
Furthermore, particles with larger impact parameters $b$ exhibit longer focal lengths, while those with smaller $b$ have shorter focal lengths.

\section{Density structure of an ALP flux focused by gravitational lensing of the Earth}
Performing the calculations described in the previous section, we generated 1000 particle trajectories for impact parameters in the range $0 \leq b \leq R_E$, with equal $R_E/1000$ steps.
To visualize the local ALP density at each spatial point, we discretized the region containing the gravitational lensing effect into discritized spatial elements.
In this analysis, we adopt a cylindrical coordinate system $(x \equiv \rho \cos \phi,\ y \equiv \rho \sin \phi,\ z)$ 
with $\rho > 0$ and $0 \leq \phi < 2\pi$, which naturally aligns with the rotational symmetry 
about the $z$-axis depicted in Fig.\ref{calc geometory}. 
This coordinate choice is most suitable for capturing the cross-sectional density distribution of the lensing object. 
In the figure, strictly speaking, the $x$-axis corresponds to the cases of $\phi = 0$ and $\phi = \pi$. 
Nevertheless, since the azimuthal origin $\phi = 0$ can be chosen arbitrarily, 
the $\pm x$ axis should be interpreted as representing an arbitrary cross-sectional plane around the $z$-axis. 
In all subsequent figures showing $x$–$z$ cross sections, this convention will be implicitly assumed.
Due to the rotational symmetry, trajectories with $b \leq 0$ can be inferred by mirroring those with $b \geq 0$ about the z-axis.
Thus, one set of calculations yields double the statistical weight, enabling us to effectively obtain 2000 trajectories over the range $-R_E \leq b \leq R_E$.
\begin{figure}[h]
\centering
\includegraphics[width=0.5\textwidth]{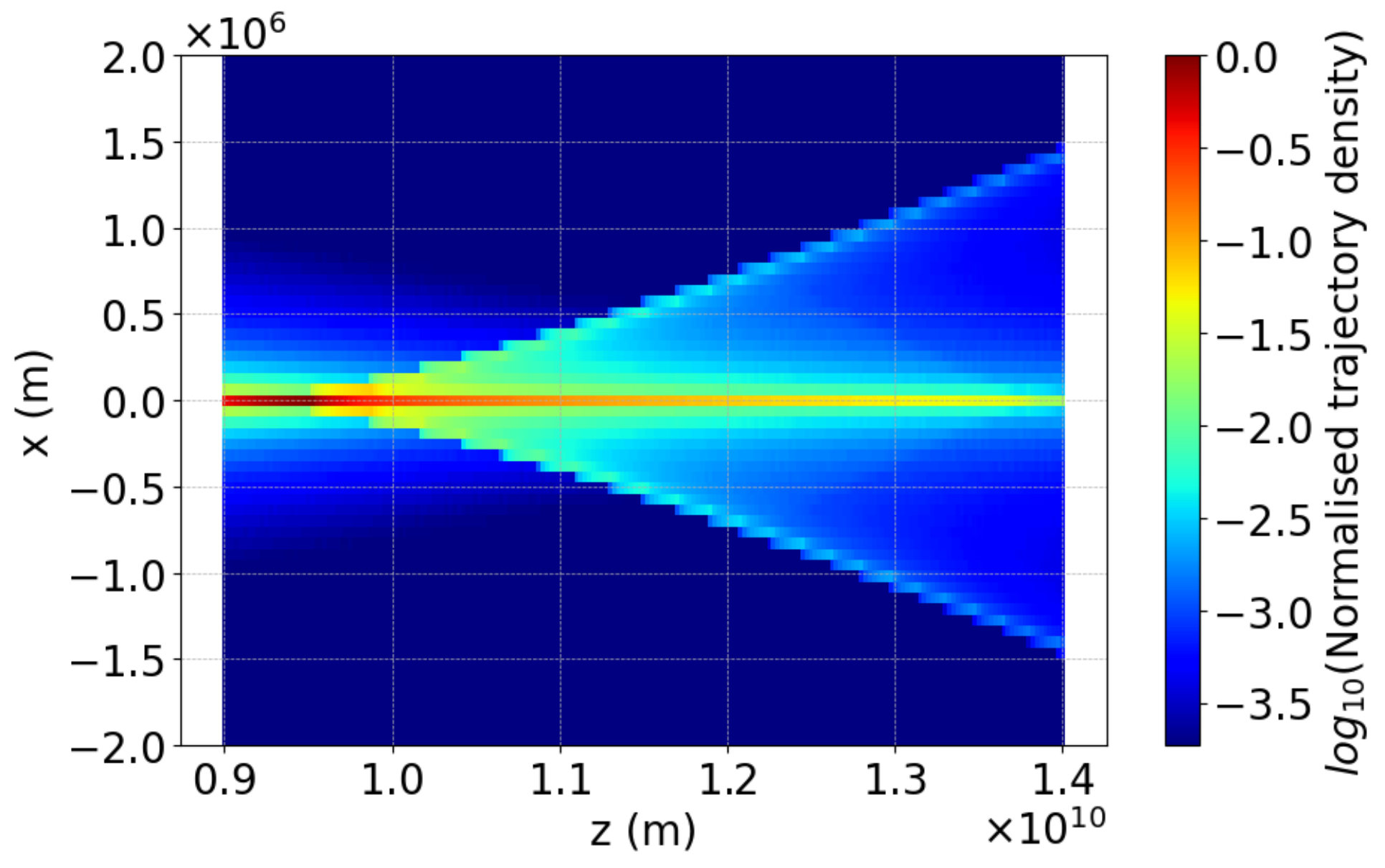}
\caption{
Normalized density profile on the x-z plane of the lensing object 
when a unidirectional dark matter flux is incident along the z-axis with the velocity of 520~km/s.
}
\label{floral lensing object}
\end{figure}

Figure \ref{floral lensing object} shows the normalized density profile on the x-z plane of the lensing object 
when a unidirectional dark matter flux is incident along the z-axis with the velocity of 520 km/s.
The element size is $2.5 \times 10^7 \,\mathrm{m}$ along the z-axis ($ (1.4 \times 10^{10} - 9.0 \times 10^9)/200 $) 
and $6.37 \times 10^4 \,\mathrm{m}$ along the x-axis ($ (R_E - (-R_E))/200$ ).
The number of trajectories in each segment is counted and the density is obtained by dividing it by the segment volume.
The density profile in the color contour are displayed after normalizing individual densities to that of 
the maxmum density segment.
As shown in Fig. \ref{floral lensing object}, the spatial distribution of the normalized ALP density 
reveals a petal-like structure.
In the surrounding regions, the density is also relatively high in the "petal" structures.
This can be understood by analogy with an optical lens: light passing through a lens converges at the focal point and then diverges beyond it, forming high and low density regions.
However, a major difference is that gravitational lensing in this context exhibits a large aberration relative to the focal length.
Aberration refers to the spreading of rays due to imperfect focusing.
As shown in Fig. \ref{forcal length dist} in Appendix of this paper, 
the focal distance increases monotonically with the impact parameter $b$, introducing a significant variation.
This dependence of aberration and focal length on $b$ explains the emergence of the petal-shaped density structure beyond the focal region.

\section{Acceptance of reflected signal photons on the detector plane}
\begin{figure}[h]
\centering
\includegraphics[width=0.5\textwidth]{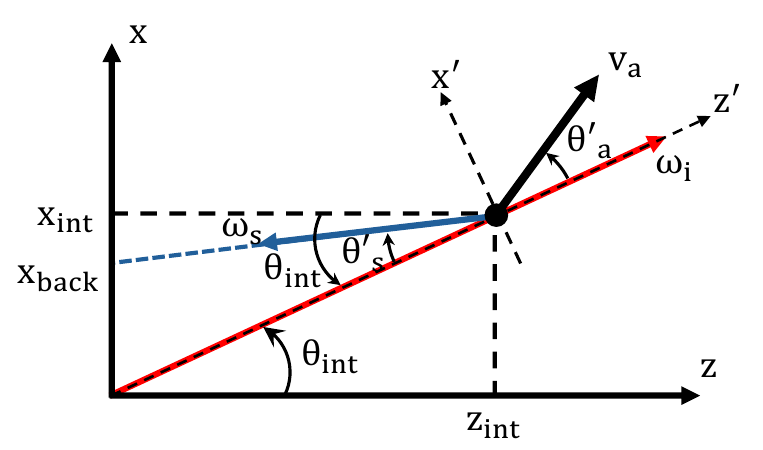}
\caption{
Geometry of the inducing field emission point, interaction point, and signal photon trajectory.
The red arrow indicates the wavevector of the inducing field.
The blue arrow and dashed line represent the emission direction of the signal photon 
and the path to the detector plane, respectively.
The black dot and arrow denote the ALP and its velocity direction, respectively.
Subscripts: “$a$” for the ALP, “$i$” for the inducing field, “$s$” for the signal photon.
The angles $\theta^\prime_a$ and $\theta^\prime_s$ represent the relative angles with respect to 
the wavevector of the inducing field.
}
\label{xback concept}
\end{figure}

In the previous work~\cite{Homma2024} 
the acceptance factor $A_{cc}$
— defined as the fraction of signal photons collected within the finite area of the detector plane — 
was roughly estimated by assuming ALPs moving at $10^{-3}c$ perpendicularly to the line of sight from the detector point, 
as a conservative approximation. In this paper, we refine the signal acceptance factor by considering ALPs traveling 
along curved trajectories.
We thus derive the expression for $A_{acc}$ based on Fig.\ref{xback concept}.
We first summarize the kinematical relations of the induced decay.
Let $x_{\mathrm{back}}$ be the transverse distance from the optical axis at which a signal photon arrives on the detector plane.
The interaction point between the inducing field and the ALP is denoted by $(x_{int}, z_{int})$.
Accordingly, the angle between the local wavevector of the inducing field and the optical axis is
$\theta_\mathrm{int} = \arctan\left(x_{int}/z_{int}\right)$.
The angle definitions of the ALP velocity direction $\theta^{'}_a$ (black arrow) and the signal emission direction $\theta^{'}_s$ (blue arrow)  at the induced decay point with respect to the local wavevector of the inducing field $\theta_\mathrm{int}$ (red arrow) are illustrated in Fig. \ref{xback concept}, where the prime symbol indicates that the local reference coordinates are based on the direction of the inducing field propagation with respect to the laboratory frame 
without the primed symbol.
In order to calculate $x_{\mathrm{back}}$ we first compute the emission angle $\theta^\prime_s$ of the signal photon at the decay point.
Let $\beta_a = v_a/c$ be its relativistic velocity ($c$ is the velocity of light) resulting in the Lorentz factor $\gamma = 1/\sqrt{1-\beta^2_a}$,
and $\omega_i$ be the energy of a single inducing photon.
Energy-momentum conservation in the local reference frame requires the following relations:
\begin{equation}\label{EqConservation}
\begin{aligned}
\text{Energy:} &\quad \gamma_a m_a = \omega_i + \omega_s \\
\text{Momentum along: } z^\prime &\quad \gamma_a m_a \beta_a \cos\theta^\prime_a = \omega_i + \omega_s \cos(\pi - \theta^\prime_s) \\
\text{Momentum along: } x^\prime &\quad \gamma_a m_a \beta_a \sin\theta^\prime_a = \omega_s \sin(\pi - \theta^\prime_s).\\
\end{aligned}
\end{equation}
From these, we can derive
\begin{equation}\label{prime cordinate sin/cos}
\cos\theta^\prime_s = \frac{\omega_i - (\omega_i + \omega_s) \beta_a \cos\theta^\prime_a}{\omega_s} \mbox{,} 
\sin\theta^\prime_s = \frac{(\omega_i + \omega_s) \beta_a \sin\theta^\prime_a}{\omega_s}.
\end{equation}
It is allowd to put $\omega_i = k m_a$ without loss of generality with a constant fraction $k$ if $k$ satisfies the following relation
\begin{equation}
\frac{1}{2}\gamma_a(1-\beta_a) \le k \le \frac{1}{2}\gamma_a(1+\beta_a)
\label{k}
\end{equation}
as the result of energy-momentum conservation in Eq.(\ref{EqConservation}).
This results in the mass independence of the decay angles because mass in $\omega_i = k m_a$ and $\omega_s = (\gamma_a-k) m_a$ are all cancelled out 
in Eq.(\ref{prime cordinate sin/cos}).
Therefore, counter-intuitively, the acceptance factor eventually becomes common for any ALP mass that we target.
As shown in Fig. \ref{xback concept}, the coordinate value of the reflected signal photon on the detector plane is expressed as
\begin{equation}
\begin{aligned}
x_{\mathrm{back}} &= x_{int} - z_{int} \tan(\theta_\mathrm{int} - \theta^{\prime}_s) \\
&= x_{int} - z_{int} \frac{\sin\theta_\mathrm{int} \cos\theta^\prime_s - \sin\theta^\prime_s \cos\theta_\mathrm{int}}{\cos\theta_\mathrm{int} \cos\theta^\prime_s + \sin\theta_\mathrm{int} \sin\theta^\prime_s}.\\
\end{aligned}
\end{equation}

In the following calculation to obtain the signal acceptance factor, a divergence angle of the inducing field, $\theta_i$, 
propagating as the spherical wave was set to $\theta_i = 3 \times 10^{-4}\,\mathrm{rad}$, and the searching distance to $5 \times 10^{9}\,\mathrm{m}$. 
Under these conditions, the distribution of the height $x_{back}$ from the center of the detector plane was obtained based on 
the numerically determined positions and velocities of individual simulated trajectories of ALPs over the searching distance.
\begin{figure}[h]
    \centering
\includegraphics[width=0.5\textwidth]{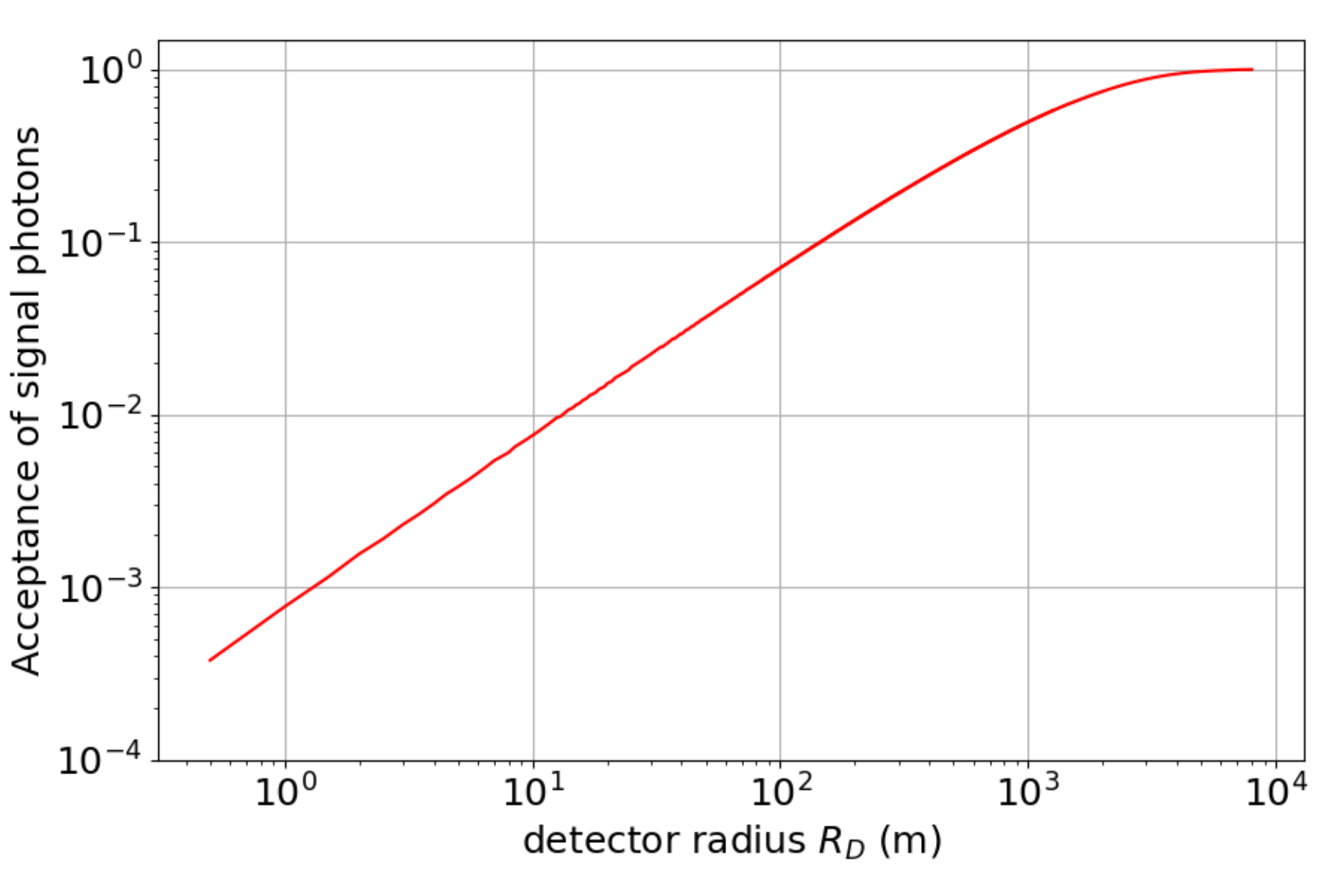}
\caption{
Acceptance factor of signal photons collectable within detector radius $R_{D}$ as a function of $R_D$. 
$R_D$ defines the integral range such that $0 \le x_{\mathrm{back}} \le R_D$ for counting the number of stimulated decay photons 
observed within the detector area.
$x_{back}$ is calculated using $\omega_i=\frac{1}{2} m_a$ with $k=\frac{1}{2}$ in Eq.(\ref{k}).
}
\label{SignalDist}
\end{figure}

Figure~\ref{SignalDist} shows the acceptance factor, $A_{cc}$, defined as the fraction of signal photons that are observed within a detector area of radius $R_D$.
The value of $R_D$ defines the integral range such that $0 \leq x_{\mathrm{back}} \leq R_D$ for counting the number of decayed photons 
observed within the detector area. The quantity $x_{\mathrm{back}}$ is calculated using $\omega_i = \tfrac{1}{2} m_a$ with $k = \tfrac{1}{2}$ in Eq.(\ref{k}).
This acceptance factor is evaluated under the condition that the detector plane is placed at a distance of $9.2 \times 10^9~\mathrm{m}$ 
from the Earth's center, illuminated by a spherical inducing wave with a beam divergence angle of $3 \times 10^{-4}~\mathrm{rad}$, 
over a search range of $5.00 \times 10^9~\mathrm{m}$.
The detector position corresponds to the location of the nearest focal point from the Earth, 
calculated using the analytical expression for the focal length in Eq.~(27) in Appendix of this paper, assuming an initial velocity of $520~\mathrm{km/s}$.
The maximum value of $x_{\mathrm{back}}$ among all the simulated trajectories with induced decays is approximately $8000~\mathrm{m}$.
Since the signal acceptance is calculated from discrete spatial points along the entire set of simulated ALP trajectories, 
the resulting $x_{\mathrm{back}}$ values are also discrete.
From Fig.~\ref{SignalDist}, it is found that approximately $\mathcal{O}(10^{-3})$ of the photons can be accepted 
using a detector of radius $R_D = 1~\mathrm{m}$. 
This surprisingly high acceptance factor over the propagation distance of $5.00 \times 10^9~\mathrm{m}$ is attributed to 
the tight collimation of the ALP flux around the focal axis and the spherical mirroring effect of on the wavefront of the inducing field.

\section{Sensitivity projection on ALP-photon coupling}
Given the focused ALP trajectories in the numerical calculation and the signal acceptance factor in the previous sections, we can move on to the calculation to evaluate the signal yield $\mathcal{Y}$ from SBR. For this calculation we move to spherical coordinates because the inducing field propagates as a spherical wave in space, which is expressed as the following equation\cite{Homma2024}:
\begin{equation}\label{SignalYield}
\mathcal{Y} = \int_{L_0/c}^{L_1/c} dt \int \rho_i\, dV_i \int \rho_a dV_i \int_0^{\Omega_i} \dv{\Gamma_0}{\Omega}\, d\Omega,
\end{equation}
where $V_i$ specifies the volume scanned by the inducing field 
while it propagates over the searching region from radial position $r=0$ to $L_1-L_0$ along polar angle $\theta=0$ assuming the location of the space probe at $z=L_0$ from the center of the Earth with the solid angle
$\Omega_i = \int^{\theta_i}_0 d\theta \sin\theta \int^{2\pi}_0 d\phi = 2\pi (1-\cos\theta_i)$
defined by the divergence angle $\theta_i$ determined by the specification of a coherent light source.
The number density of the inducing field $\rho_i$ is parametrized as
$\rho_i(r) = \frac{N_i}{c\tau r^2 \Omega_i}$ as a function of $r$
reflecting the situation that the pulse center position moves along the r-axis
with the surface area of $r^2 \Omega_i$ at $r$.

\begin{figure}[h]
\centering
\includegraphics[width=0.4\textwidth]{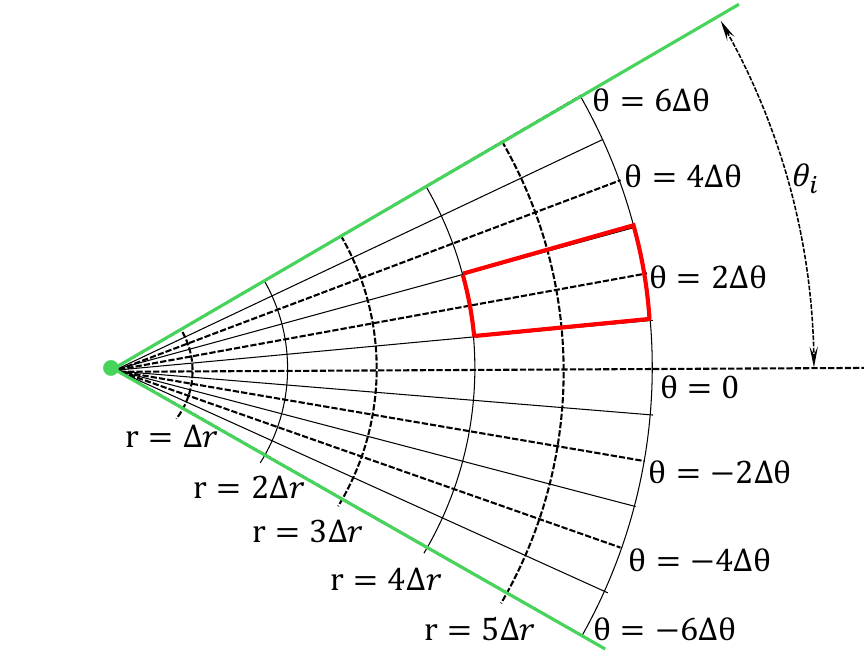}
\caption{Concept of segments in the propagation volume of an inducing field.
The propagation region of the inducing field, shown inside of two green line, is finely divided into segments with intervals $\Delta r$ and $\Delta\theta$.
This shows a configuration when the divergence angle of the inducing beam is divided into 12 equal angle range $\Delta\theta$.
The center of each segment takes values at $(r, \, \theta) = (2n \, \Delta r, \, 2m \, \Delta \theta)  \, (\text{with } n,\,m \geq 0)$.
As an example, the segment outlined in red has its center at $(r, \, \theta) = (4\Delta r, \, 2\Delta \theta)$, 
and its boundaries are defined by the ranges $r - \Delta r \leq r \leq r + \Delta r$ and $\theta - \Delta \theta \leq \theta \leq \theta + \Delta \theta$.}
\label{Segmentation}
\end{figure}

In order to evaluate $\rho_a$ from the numerically generated trajectories,  we divide the region of the focused ALP flux into segments in sperical coordinates instead of x-z coordinates as indicated in Fig.\ref{Segmentation}. Accordingly the normalized density profile on the x-z plane is recalculated in sperical coordinates.
We introduce the initial ALP number density sourced by the S1 stream $\rho_{in}$ with
Eq.(\ref{eq_frho}) as follows 
\begin{equation}\label{rho in}
    \rho_{in}=\rho_0 f_\rho/m_a.
\end{equation}
We then introduce ALP densities $\rho_a(r, \, \theta)$ in individual segment centers at $(r, \,  \theta)$ as
\begin{equation}\label{ALP concentrate density}
\rho_a(r, \, \theta) = \rho_{in} \mathtt{M} \frac{n(r,\theta)}{n_{\mathtt{M}}},
\end{equation}
where $\mathtt{M}$ is the maximum density enhancement factor, in other words, the maximum magnification factor
along the incident axis of the ALP flux and $n_\mathtt{M}$ is the number density of trajectories 
in the segment of the maximum magnification. 
The definition of magnification is provided in Appendix of this paper which summarizes the content of Ref.~\cite{Przeau_2015}.
Given the volume of a segment for the inducing field as $V_i(r,\theta)$ resulting from the rotation of the differential solid angle about the axis with $\theta = 0$ in spherical coordinates, which is defined as
\begin{equation}
V_i(r,\theta) = \int_{0}^{2\pi}d\phi^\prime \int_{r-\Delta r}^{r+\Delta r}dr^\prime \int_{\theta-\Delta \theta}^{\theta+\Delta \theta}d\theta^\prime\, r^{\prime2}\sin\theta^\prime,
\end{equation}
the number density of trajectories $n(r,\theta)$ is expressed as
\begin{equation}
  n(r,\theta)=\frac{\mathcal{N}_\mathrm{V}(r,\theta)}{V_i(r,\theta)}
\end{equation}
where $\mathcal{N}_\mathrm{V}(r,\theta)$ is the number of trajectories in the inducing volume element $V_i(r,\theta)$, corresponding to $\mathcal{N}_\mathrm{V}(r,\theta) \equiv \mathcal{N}_\mathrm{S}(r,\theta) 2\pi r \sin\theta$ with the number of trajectories $\mathcal{N}_\mathrm{S}(r,\theta)$ counted in the $r-\theta$ plane in Fig.\ref{Segmentation}.  
 
The differential decay rate for ALP decaying into two photons is given as follows \cite{Homma2024}:
\begin{equation}
    \dv{\Gamma_0}{\Omega} \equiv \frac{1}{128\pi^2 m_a} \left( \frac{g}{M} \right)^2 \omega_i^4 \cos^2 \Phi,
\end{equation}
where $m_a$ is the ALP mass, $g/M$ is the ALP–photon coupling constant, 
$\omega_i$ is the energy of the inducing photon, 
$\Phi$ is a relative angle between the direction of the electric field component of the linear polarization 
of a decayed photon with respect to that of the magnetic field component of the linearly polarized inducing photons,
and we neglected the effect of the Lorentz boost of moving ALP due to the non-relativistic nature.
Assuming $\Phi=0$ with the signal polarization selection in the detector part~\cite{Homma2024},
the integral form in Eq. (\ref{SignalYield}) can thus be rewritten as an equivalent discrete summation as follows
\begin{eqnarray}\label{discreat coupling formula}
\mathcal{Y} \equiv \frac{\omega_i^4 (L_1 - L_0)(1 - \cos \theta_i)}{64 \pi m_a c} \times \\ \nonumber
\left( \sum_{r,\,\theta} \rho_a(r,\theta)V_i(r,\theta)N_i(r,\theta) \right) \left( \frac{g}{M} \right)^2,
\end{eqnarray}
where the summation term $\sum_{(r,\theta)} \rho_a(r,\theta)V_i(r,\theta) N_i(r,\theta)$ enables the incorporation of the local overlap between the inducing field and the ALP density with
\begin{eqnarray}
N_i(r,\theta) = \frac{\int_{0}^{2\pi}d\phi^\prime \int_{\theta-\Delta \theta}^{\theta+\Delta \theta}d\theta^\prime\, \sin\theta^\prime}{\Omega_i} \frac{E_i}{\omega_i}\\ \nonumber
= \frac{\int_{0}^{2\pi}d\phi^\prime\int_{\theta-\Delta \theta}^{\theta+\Delta \theta}d\theta^\prime\, \sin\theta^\prime}{2\pi(1-\cos\theta_i)} \frac{E_i}{\omega_i},
\end{eqnarray}
where energy of a single laser pulse $E_i$, solid angel $\Omega_i$, and divergence angle  $\theta_i$  of the inducing field.

In the single photon counting, we assume a photomultiplier tube (PMT) for the laser frequency. 
Although thermal noise in PMT may remain as a residual background, 
the primary source of background around the detector part in space is charged cosmic rays. 
At distances of approximately $\mathcal{O}(10^{10})\,\mathrm{m}$ from the Earth, the dominant components are galactic cosmic rays.
Among these, particles with energies above 1 GeV can enter the PMT photocathode at a flux of approximately $1\,\mathrm{Hz/cm^2/sr}$.
Assuming the photocathode area is on the order of $\mathcal{O}(1)\,\mathrm{cm^2}$, the expected rate of background photoelectron events due to cosmic rays is
$1\,\mathrm{Hz/cm^2/sr} \times \mathcal{O}(1)\,\mathrm{cm^2} \times 4\pi\,\mathrm{sr} = \mathcal{O}(10)\,\mathrm{Hz}$.
In this work, we therefore estimate the background rate as $R_{\mathrm{bkg}} = 10\,\mathrm{Hz}$.
 
Based on Eq.(\ref{discreat coupling formula}),
the observed number of signal photons via stimulated backward reflection (SBR) is expected to be
\begin{equation}
\mathcal{Y}_{obs} = \mathcal{Y} T f \epsilon_D A_{cc}
\label{Yobs}
\end{equation}
with measurement time $T$, repetition rate of the inducing laser pulse $f$, photon detection efficienty $\epsilon_D$ and the acceptance factor $A_{cc}$.
By equating the $5\sigma$ level of background fluctuation to the above observed signal yield,
the upper limit of the coupling $g/M$ under the assumption of a null result is expressed as
\begin{widetext}
\begin{equation}
\frac{g}{M} = \sqrt{ \frac{ \delta N_{bkg} }{ f T \epsilon_D A_{cc} \frac{ \omega_i^4 (L_1 - L_0)(1 - \cos \theta_i) }{ 64 \pi m_a c } \left( \sum_{(r,\theta)} \rho_a(r,\theta)V(r,\theta)N_i(r,\theta) \right) } }
\end{equation}
\end{widetext}
with $\delta N_{bkg} \equiv 5 \sqrt{2 T R_{bkg}}$ \cite{Homma2024}.

\begin{figure}[b]
\includegraphics[width=0.5\textwidth]{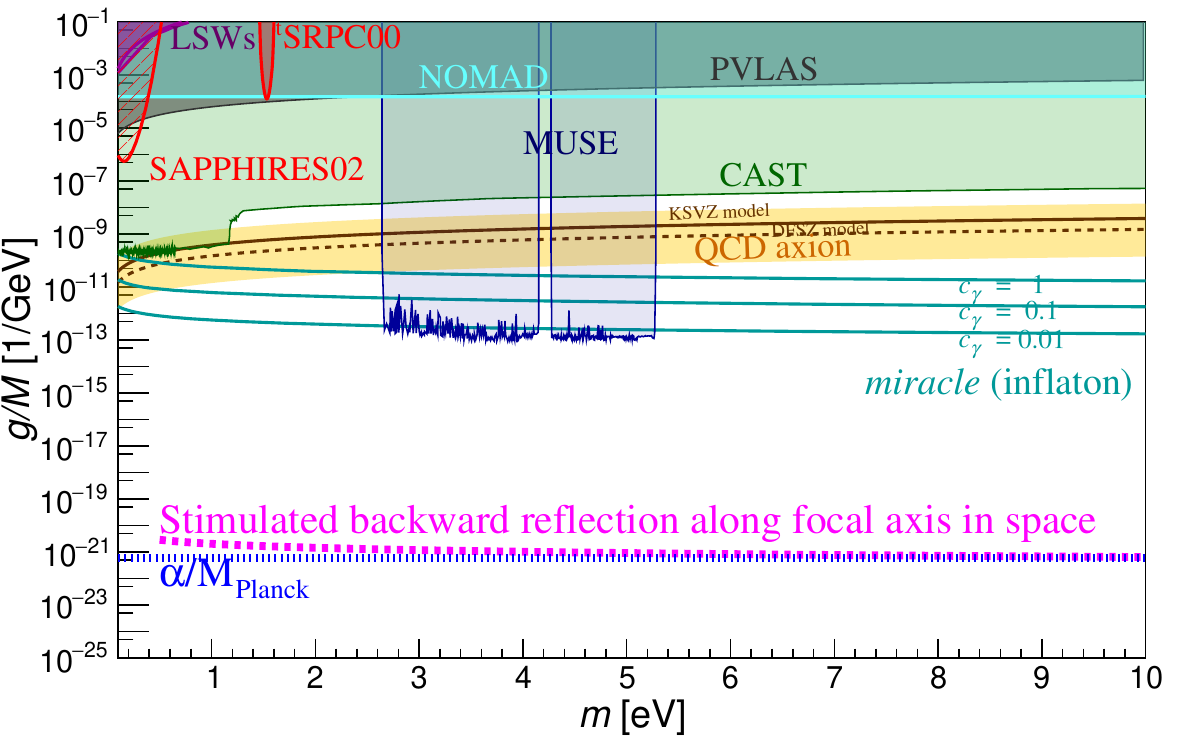}
\caption{
Sensitivity projection with the stimulated backward reflection (SBR) method applied to the focused ALP flux
by the Earth-lens effect, assuming the S1 stream as a distant ALP source.
The megenta curve shows the expected upper limit based on this Earth-lens telescope concept with parameters
summarized in Tab.\ref{Tab1}. The other details are explained in the main text.
}
\label{Projection}
\end{figure}

The parameters used for the sensitivity estimation are summarized in Tab.\ref{Tab1}.
Under these conditions, the projected sensitivity using the SBR method is shown in Fig. \ref{Projection}.
To contextualize the refined theoretical predictions presented in this work, we compare them with a representative set of experimental constraints widely referenced in the literature. The solid brown curve and surrounding yellow band illustrate the expected parameter space from the benchmark QCD axion model, specifically the KSVZ model\cite{KSVZ} with $E/N = 0$ and $0.07 < |E/N - 1.95| < 7$. The dashed brown curve denotes the prediction from the DFSZ model\cite{DFSZ1,DFSZ2} with $E/N = 8/3$. Cyan curves correspond to the ALP Miracle model relevant to inflation\cite{ALP_miracle2018}, plotted for values of $c_\gamma = 1$, 0.1, and 0.01.
Shaded regions in the plot indicate existing exclusion limits from various experimental searches: vacuum magnetic birefringence (PVLAS\cite{PVLAS2020}, gray), Light-Shining-through-a-Wall experiments (ALPS\cite{ALPS2010} and OSQAR\cite{OSQAR2015}, purple), penetrating scalar particle searches at the eV scale using the SPS neutrino beam (NOMAD\cite{NOMAD2000}, light cyan), the optical MUSE-faint survey\cite{MUSE2021} (blue), the solar helioscope experiment CAST\cite{CAST:2017uph} (green), and two beam stimulated resonant photon collider (SRPC) experiments—SAPPHIRES02\cite{SAPPHIRES02} in quasi-parallel collision geometry (hatched red) and three beam SRPC, $^ t$SRPC00\cite{tSRPC00} with the fixed incident angle setup (solid red).
 
\begin{table*}[]
\caption{Parameters used for the estimation of the sensitivity curve in Fig.\ref{Projection} 
assuming the laser-driven frequency range, that is, around the eV range which can cover 
the ALP mass range via $\omega_i=m_a/2$ for the inducing field.}
\begin{center}
\begin{tabular}{ l r r } \hline
    Parameter&Value&Units \\ \hline
    Measurement time $T$&$1.0 \times 10^6$&sec\\ 
    Divergence angle of the inducing laser $\theta_i$&$ 3 \times 10^{-4}$&rad  \\ 
    Emission point of the inducing field $L_0$&$ 9.2 \times 10^9$&m\\ 
    End point of the search region $L_1 $&$ 1.4 \times 10^{10}$&m\\ 
    Laser pulse energy $E_i$&$1.0$&J \\ 
    The number of inducing photons $N_i(m_a)$&$E_i/\omega_i = E_i/(m_a/2)$&1 \\
    Laser repetition rate $f$&$1.0$&Hz \\ 
    Radius of the signal detector surface $R_D$&$1.0$&m \\ 
    Detection efficiency of the signal photons $\epsilon_D$
    &$0.1$&1 \\ 
    Noise rate of the signal detector $R_{bkg}$&$10\,$&Hz \\ 
    S1+Halo ALP energy density$\rho_0$&0.5&GeV/$\mathrm{cm^3}$\\
    S1 stream fraction in $\rho_0$&0.002&1\\
    Incident ALP number density from S1, $\rho_{in}$ in Eq.~(\ref{rho in})&$ 1.0 \times 10^{12}/m_a$[eV]&m${}^{-3}$ \\ 
    Maximum magnification $\mathtt{M}$&$5.63 \times 10^9$&1 \\ \hline
\end{tabular}
\end{center}
\label{Tab1}
\end{table*}
 
\section{Conclusion}
We envisioned a space observatory optimized for detecting distant ALP sources, considering the S1 stream 
as a concrete example of a remote source. The proposed telescope consists of the Earth's gravitational lens and 
a collector that captures stimulated backward reflections (SBRs) from remote ALP decay points 
illuminated by a spherically propagating laser field.
First, we revealed the density structure of the focused ALP flux based on numerically simulated trajectories. 
Then, we established a computational scheme to evaluate the collection of reflected signal photons using the SBR method. 
Based on this telescope concept, we refined the sensitivity estimation in the eV mass range. 
This more accurate evaluation, which accounts for ALPs moving along curved trajectories, shows that 
the sensitivity can reach 
$\mathcal{O}(10^{-22}) \, \mathrm{GeV}^{-1},$
consistent with previously published results~\cite{Homma2024} based on the rough approximations with similar parameters.
Since the sensitivity to generic couplings is expected to reach values as small as 
$g/M = \alpha/M_{\mathrm{Planck}},$
this proposal demonstrates the feasibility of a novel space observatory capable of probing more distant ALP sources 
well beyond $ \mathcal{O}(10) \, \mathrm{kpc},$
assuming, as one illustrative case, the DFSZ axion-photon coupling upper limit of 
$g/M \sim 10^{-11} \, \mathrm{GeV}^{-1}$ in the eV mass range.

\section{Future prospects and discussions}
The existence of gravitationally bound axion or ALP configurations at distances of $10$–$100~\mathrm{kpc}$ from the Solar System is not only theoretically natural but under active observational investigation. 
The key points are summarized as follows:

\begin{itemize}
    \item Axion miniclusters and axion stars are expected to be distributed throughout the Milky Way halo, including the $10$–$100~\mathrm{kpc}$ shell, especially in post-inflationary scenarios. Numerical studies show that a significant fraction of miniclusters survive stellar tidal disruption even at $r \gtrsim 10~\mathrm{kpc}$~\cite{Tinyakov2021}.
    
    \item These miniclusters can undergo Bose condensation in their cores to form \emph{axion stars}, 
which may collapse and emit relativistic axions through processes such as ``bosenovae'' or gradual evaporation 
via self-interactions~\cite{Levkov2017, Fox2023}.
    
    \item Known magnetars in the Large Magellanic Cloud (LMC), such as SGR~0526$-$66 and J0456$-$69, located at $\sim 50~\mathrm{kpc}$, serve as realistic candidates for ALP sources through photon-ALP conversion in strong magnetic fields~\cite{Fortin2022}.
    
    \item The historical core-collapse supernova SN~1987A in the LMC has already been used to constrain ALP emission via neutrino duration and energy considerations~\cite{Raffelt1988}.
\end{itemize}

From a detection perspective, ALPs emitted from above mentioned sources may reach the Earth 
with a flux detectable by the proposed Earth-lens telescope.
The proposed detection principle is applicable to inducing fields of any frequency. 
Currently available coherent electromagnetic sources on the ground can cover the range of $10^{-6}$ 
to $10$ eV, considering the physical dimensions of the sources. 
For instance, in the $\mu$eV mass range, klystrons are commercially available. 
However, the available bandwidth, physical weight, and power consumption would pose 
realistic challenges for installing such sources on a space probe.

In our calculation, we have assumed a constant density for the Earth and 
have neglected the effect of its slow rotation.
Regarding the former simplification, although we plan to eventually incorporate the Earth's 
internal inhomogeneity with future upgrades in computational resources, 
reference~\cite{Przeau_2015} has already evaluated the effect and found that its impact 
on the focal length and magnification is not significant.
On the other hand, the latter simplification can be justified by the following discussion.
Based on the weak–field analysis of a rigidly‐rotating homogeneous sphere  
by Sereno (2003, Eq.~(21)–(22))~\cite{Sereno2003}, the two transverse components
of the deflection vector for a test particle that crosses the
\emph{equatorial plane} on a straight line ($x=b,\;y=0$ at closest
approach) read, to first order in spin $J$~\footnote{%
The (scalar) spin parameter is the magnitude of the angular--momentum vector
\[
  J\;\equiv\;|\mathbf J|
    =\int_{\text{body}}\rho(\mathbf r)\,
      \bigl(\mathbf r\times\mathbf v\bigr)\,d^{3}r ,
\]
which for a rigidly rotating body reduces to $J=I\Omega$ with
$\,I=\int\rho(r)\,r_{\perp}^{2}\,d^{3}r$ the moment of inertia and
$\Omega$ the uniform angular velocity.
For a homogeneous sphere $I=\tfrac{2}{5}MR^{2}$, hence
$J=\tfrac{2}{5}MR^{2}\Omega$.
Sereno works in geometrised units $G=c=1$;
We restore $G,c$ explicitly in Eqs.~\eqref{eq:alpha_t_spin}--\eqref{eq:defl_total}.
$U=L/(cM_{\rm tot}R_{\!E})$ is a dimensionless spin parameter.
We display only the terms that survive for $y=0$:
the frame–dragging contribution is then \emph{purely} tangential.}  
\begin{align}
\alpha_{r}^{\,(1\mathrm{PN})} &= \frac{1}{b}, \label{eq:alpha_r_mass}\\
\alpha_{t}^{\,(J)} &= \pm\frac{2U}{b^{2}}
                   = \pm\frac{4GJ}{b^{2}c^{3}v}\;, \label{eq:alpha_t_spin}
\end{align}
where $r$ and $t$ denote, respectively, the radial (lens–centred) and
tangential directions in the lens plane and the upper (lower) sign corresponds
to prograde (retrograde) motion.
Introducing the physical velocity
$\beta\equiv v/c$ and restoring $G,c$, the total deflection angle becomes
\begin{equation}
\delta\alpha \;=\;
\frac{2GM}{bv^{2}}\Bigl(1+\beta^{2}\Bigr)
\;\pm\;
\frac{4GJ}{b^{2}c^{2}v}\;.
\label{eq:defl_total}
\end{equation}
For the numerical estimate for the Earth with
$M_E=5.97\times10^{24}\,\mathrm{kg}$,\;
$J_E=7.07\times10^{33}\,\mathrm{kg\,m^{2}\,s^{-1}}$,\;
$b\simeq R_E=6.37\times10^{6}\,\mathrm{m}$,\;
$v=\beta c\;( \beta=10^{-3})$,  
Eqs.~\eqref{eq:alpha_r_mass}–\eqref{eq:defl_total} give
\begin{align*}
\delta\alpha_{\mathrm{mass}}
  &\simeq \frac{2GM_E}{bv^{2}}
           \bigl(1+\beta^{2}\bigr)
         \;=\;1.4\times10^{-3}\;\mathrm{rad},\\[4pt]
\delta\alpha_{\mathrm{spin}}
  &\simeq \pm\frac{4GJ_E}{b^{2}c^{2}v}
         \;=\;\pm1.7\times10^{-12}\;\mathrm{rad},
\end{align*}
so that
$
|\delta\alpha_{\mathrm{spin}}|/\delta\alpha_{\mathrm{mass}}
\simeq1.2\times10^{-9}.
$
Even for the slow ALP velocity $\beta\simeq10^{-3}$ the
Lense–Thirring shear is therefore nine orders of magnitude below the
static Schwarz\-schild bending, which can be safely ignored in the proposed setup.
Since the Earth's spin effect is negligiblly small, the incident anle dependence of
the focal position is also negligiblly small given the angle coverage of the inducing laser,
$\theta_i = \mathcal{O}(10^{-4})$~rad.

The effect of the Earth's magnetic field via direct coupling to ALPs is expected to be 
negligibly small, due to the low conversion probability
$P_{a\rightarrow\gamma} \sim (g/M \cdot B \cdot L / 2)^2 \sim 10^{-9}$
for a coupling strength $g/M=10^{-7}$~GeV${}^{-1}$, with $B \sim 10^{-4}$~T and $L\sim R_E$.
Unless the conversion probability approaches unity, the loss of ALP flux cannot become 
a practical problem. On the other hand, the effect of charged particles near the detection region 
in the Van Allen belts may constitute a background source for the photon counter 
if the space probe is located within the belts. However, since the optimum detection position,
$L_0=9.2 \times 10^9$~m from the Earth's center 
is much further than the maximum extent of the belts, $\mathcal{O}(10^8)$~m,
this effect is also expected to be strongly suppressed.

\section{Acknowledgement}
T. Nakamura thanks Haruhiko Nishizaki for useful discussions on the efficiency and
generality of the source code used for numerical computations and analysis, as well as on general relativity.
K. Homma acknowledges the support of the Collaborative Research Program of the Institute
for Chemical Research at Kyoto University (Grant Nos. 2024–95 and 2025-100), JSPS
Core-to-Core Program (grant number: JPJSCCA20230003), and Grants-in-Aid for Scientific
Research (Nos. 21H04474 and 24KK0068) from the Ministry of Education, Culture, Sports,
Science and Technology (MEXT) of Japan.

\section{Appendix}
\subsection{Derivation of magnification $\mathtt{M}$}
\begin{figure}[h]
\centering
\includegraphics[width=0.50\textwidth]{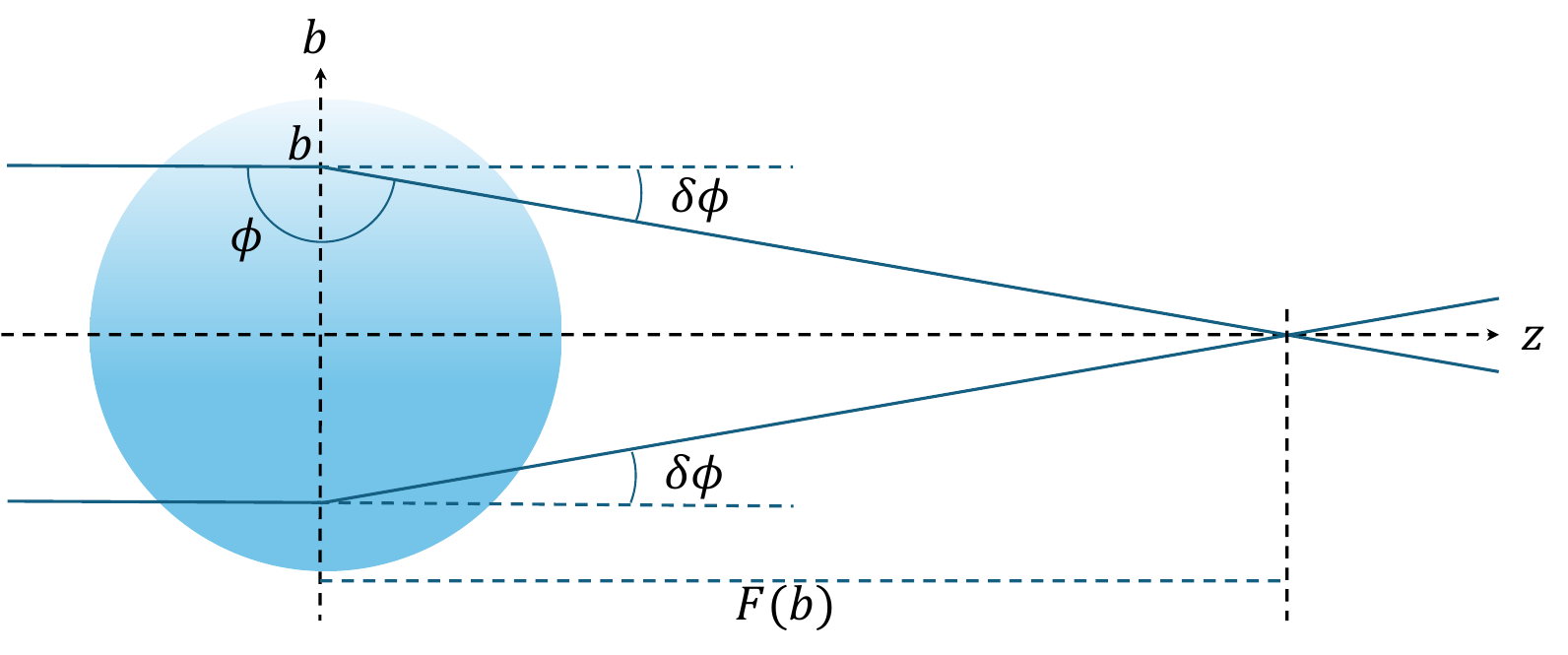}
\caption{
Geometric relation between the focal distance and the impact parameter.
The flux passing through the Earth is deflected by an angle $\delta\phi$, reaching a focal distance F(b).
}
\label{lensing geometory}
\end{figure}

The expression for $\mathtt{M}$, used to compute $\mathtt{M}_{\mathrm{max}}$ in Eq.(\ref{ALP concentrate density}), and its derivation are given below.
From the geodesic calculation of massive particles passing through the Earth on the Schwarzschild metric\footnote{metric for massive particle passing the Earth is described as~\cite{Przeau_2015}: $\left(1+\frac{v^2}{c^2}\right)e^{-2\Phi/c^2}-1-\frac{v^2 b^2}{c^2 r^2} =\left[1-\frac{r_s \mathrm{M}_{eff}(r)}{r M_E}\right]^{-1}\left(\dv{r}{\phi}\right)^2\left(\frac{v^2 b^2}{c^2 r^2}\right)$ where $\mathrm{M}_{eff}(r)$ denotes the mass enclosed within radius $r \leq R_E$ inside the Earth.}
the azimuthal angle near the gravitational lens is expressed as~\cite{Przeau_2015}
\begin{equation}\label{deflection angle}
    \begin{aligned}
        \phi &= \phi^\prime_{\mathrm{M}_{eff}} + \frac{1}{2}\arccos\left[\frac{u^2-\frac{1}{2}\alpha}{\sqrt{\frac{1}{4}\alpha^2-\beta}}\right]
    \end{aligned}
\end{equation}
with the velocity of light $c$ and $u\equiv1/r$, where $\phi^\prime_{\mathrm{M}_{eff}} ,\,\alpha ,\,\beta,\,\xi$ are defined as follows:
\begin{equation}
    \begin{aligned}
        \phi^\prime_{\mathrm{M}_{eff}} &= \phi^\prime_{E}
        + \arccos\left[\frac{\xi b^2/R_E -1}{\sqrt{1 + \xi^2 b^2}}\right]
        - \frac{1}{2}\arccos\left[\frac{2/(\alpha R_E^2) - 1}{\sqrt{1-4\beta/\alpha^2}}\right]\\
        \phi^\prime_{E} &= \arccos(1+\xi^2 b^2)^{-1/2}\\
        \alpha &= \frac{1}{b^2} + \frac{3}{2}\frac{r_S}{R_E b^2}\left(1 + \frac{c^2}{v^2}\right) + \frac{r_S}{R_E^3}\\
        \beta &= \frac{r_S}{2R_E^3 b^2}\left(3+\frac{c^2}{v^2}\right)\\
        \xi &= \frac{v^2}{G M_E},
    \end{aligned}
\end{equation}
where $r_S=2GM_E/c^2$ is the Schwarzschild radius of the Earth, $b$ is the ALP’s impact parameter relative to the Earth, 
and $v$ is the incident velocity.
The deflection angle $\delta\phi$ of the lens is then given by
\begin{equation}
    \delta \phi = \pi - \phi.
\end{equation}
From the geometry involving Earth’s center, the focal point of the lens, and the point of ALP incidence (as shown in Fig. \ref{lensing geometory}), the focal distance $F(b)$ for a ALP incident with impact parameter b is given by
\begin{equation}\label{analytic forcal length}
F(b) = \frac{b}{\delta \phi}.
\end{equation}

\begin{figure}[]
\centering
\includegraphics[width=0.45\textwidth]{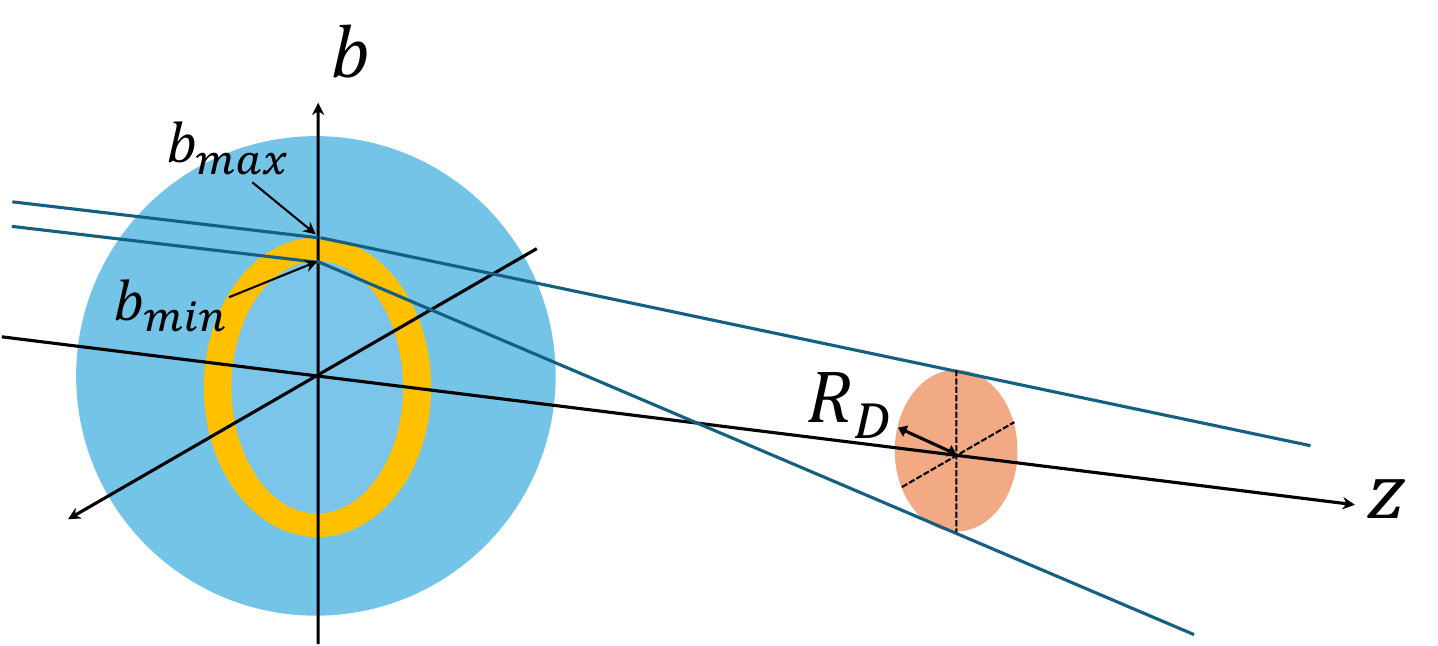}
\caption{
Conceptual diagram to define magnification $\mathtt{M}$.
Shown are the detector placed along the incident axis (orange disk), the range of impact parameters for the flux passing through the disk (yellow annulus), and the trajectories corresponding to $b_{\mathrm{max}}$ and $b_{\mathrm{min}}$ (solid navy lines).
}
\label{Magnification concept}
\end{figure}

When a detector with radius $R_D$ is placed at a distance $F(\bar{b})$ along the incident axis (the z-axis), the range of impact parameters for the flux passing through the detector is defined as $b_{\mathrm{min}} \leq b \leq b_{\mathrm{max}}$.
The magnification $\mathtt{M} = \mathtt{M}(\bar{b})$ is expressed as the ratio between the area of the annular section of Earth’s cross-section corresponding to $b_{\mathrm{min}} \leq b \leq b_{\mathrm{max}}$ and the area of the detector. Specifically, it is given by~\cite{Przeau_2015}
\begin{equation}
\mathtt{M} = \mathtt{M}(\bar{b}) = \frac{b_{\mathrm{max}}^2 - b_{\mathrm{min}}^2}{R_D^2}.
\end{equation}
The geometric relation for this configuration is illustrated in Fig.~\ref{Magnification concept}.

\subsection{Comparison of numerical calculation with the analytical prediction}
\begin{figure}[]
\centering
\includegraphics[width=0.45\textwidth]{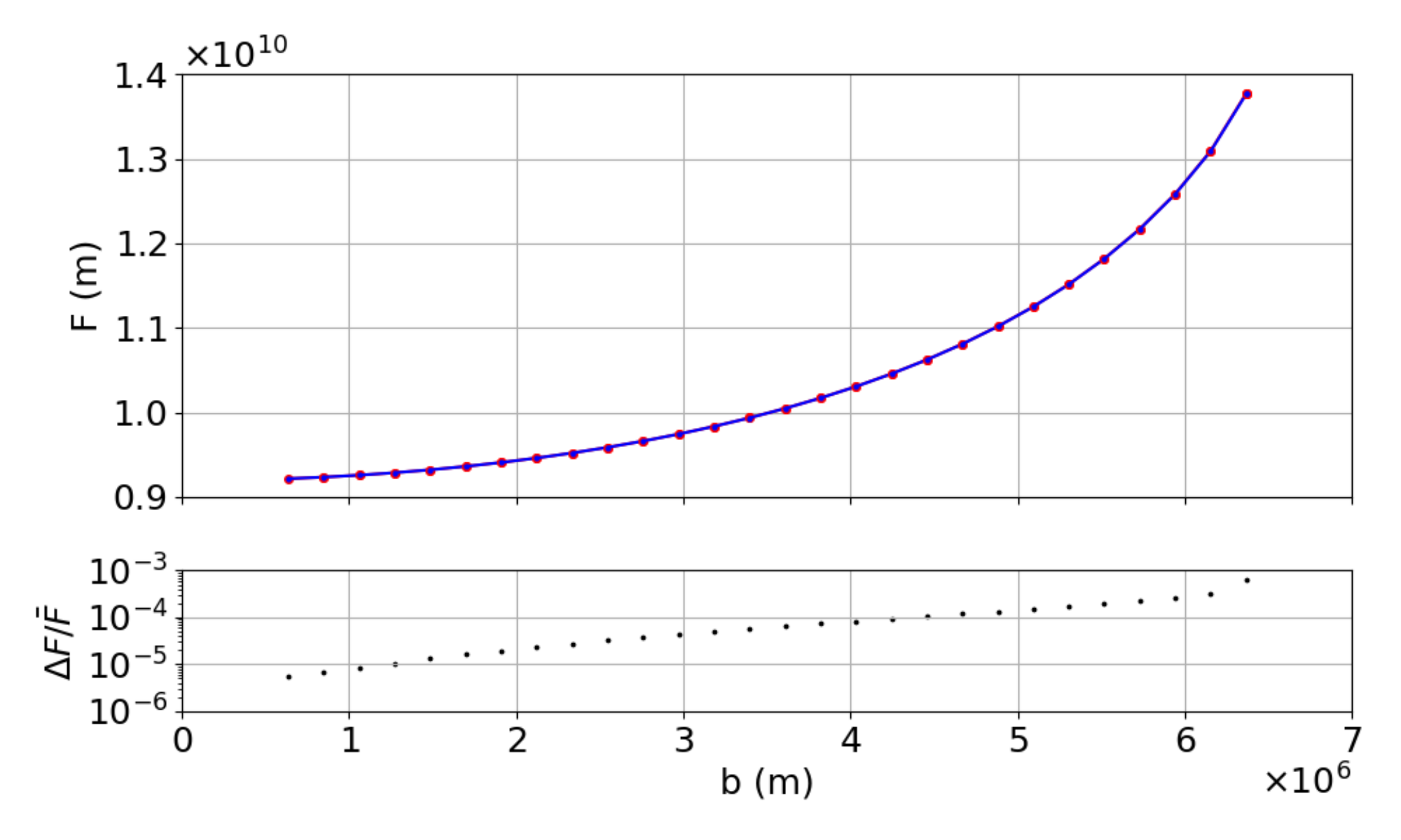}
\caption{Dependence of the focal length $F$ on the impact parameter $b$.
The red dots in the upper plot show the focal length obtained from the numerical calculation, 
while the blue line shows the analytical computation given by Eq.(\ref{analytic forcal length}).
The black dots in the lower plot show the relative difference between numerical and analytical results: 
$\Delta F/ \bar{F} = (F_{\mathrm{num}}(b) - F_{\mathrm{analy}}(b))/\bar{F}$
with $\bar{F}(b) = \frac{F_{\mathrm{num}}(b) + F_{\mathrm{analy}}(b)}{2}$.
}
\label{forcal length dist}
\end{figure}

Based on the analytical expression presented in the previous section, we calculate the focal length.
The analytical formula used to compute the focal distance is given in Eq.(\ref{analytic forcal length}).
Using this expression, we perform the calculation for an incident velocity of $v = 520\,\mathrm{km/s}$ and compare the results with those obtained from numerical integration.
The comparison is shown in Fig. \ref{forcal length dist}, demonstrating that the numerical results reproduce the analytical solution with high accuracy.

\newpage
\bibliography{cite}

\end{document}